\documentclass[aps,prd,onecolumn,groupedaddress,showpacs,nofootinbib,amssymb]{revtex4}
\usepackage[dvips]{graphicx}
\usepackage{amssymb}
\usepackage{diagbox, float}
\usepackage{amsmath}
\usepackage{graphicx,color}
\usepackage{amsfonts}
\usepackage{bm}
\usepackage{cancel}
\usepackage{comment}

\newcommand\be{\begin{equation}}
\newcommand\ee{\end{equation}}

\newcommand{\bea}{\begin{eqnarray}}
\newcommand{\eea}{\end{eqnarray}}

\allowdisplaybreaks[4]

\begin{document}

\title{Evolution of perturbations in the model of Tsallis holographic dark energy}
\author{Artyom V. Astashenok}
\email{aastashenok@kantiana.ru}
\author{Alexander S. Tepliakov}
\affiliation{I. Kant Baltic Federal University\\
236041, Kaliningrad, Russia\\}%

\tolerance=5000

\begin{abstract}

We investigated evolution of metric and density perturbations for Tsallis model of holographic dark energy with energy density $\sim L^{2\gamma - 4}$, where $L$ is length of event horizon or inverse Hubble parameter and $\gamma$ is parameter of non-additivity close to $1$. Because holographic dark energy is not an ordinary cosmological fluid but a phenomena caused by boundaries of the universe, the ordinary analysis for perturbations is not suitable. One needs to consider perturbations of the future event horizon. For realistic values of parameters it was discovered that perturbations of dark energy don't grow infinitely but vanish or freeze. We also considered the case of realistic interaction between holographic dark energy and matter and showed that in this case perturbations also can asymptotically freeze with time. 

\end{abstract}


\maketitle
\section{Introduction}
\label{sec1}

In 1998, it was discovered that the universe expands with acceleration
\cite{1}, \cite{2}. This required a new component in the universe, called
dark energy, which has low density and only interacted with gravity. The
simplest way to model dark energy was to use Einstein's $\Lambda $ as a
cosmological constant or vacuum energy. This resulted to the
$\Lambda $ CDM model
\cite{LCDM-1,LCDM-2,LCDM-3,LCDM-4,LCDM-5,LCDM-6,LCDM-7}, which is the standard
cosmological model today. Despite of the good agreements with observational
data this model has some problems on the theoretical level. Firstly, there
is a problem of smallness of cosmological constant. From considerations
based on quantum field theory it follows that vacuum energy should be approximate
to the Planck value which dramatically contradicts to the observations.
Also the cosmic coincidence problem appears: why is current matter density
is close to the value of vacuum density? The answer to this question is
unclear in frames of standard cosmology.

There are many other ways to explain cosmic acceleration. Various models
in which dark energy is some scalar field are considered (see
\cite{Quint} and references therein). The simplest scalar-field scenario
without ghosts and instabilities is quintessence. Quintessence is described
by a scalar field with positive kinetic term minimally coupled to gravity.
A slowly varying field along an appropriate potential can lead to the acceleration
of the Universe.

The more exotic phantom model was firstly investigated in ref.~\cite{Caldwell}.
Although this model does not contradict the cosmological tests based on
observational data, the phantom field is unstable \cite{Carrol} because
the violation of all energy conditions occurs. Moreover, phantom energy
model with $w=\text{const}$ has Big Rip future singularity
\cite{Frampton,S,BR,Nojiri-2,Faraoni,GD,Elizalde,Sami,Csaki,Wu,NP,Stefanic,Chimento,Dabrowski,Godlowski,Sola,Nojiri}: the scale factor becomes infinite at a finite
time.

Another possibility is various modifications of gravity
\cite{reviews1,reviews2,reviews3,reviews4,book,reviews5,Harko}. Modified
gravity models assume that the universe is accelerating due to the deviations
of real gravity from general relativity on cosmological scales. Any theory
of modified gravity should be tested on astrophysical level also because
one can hope that strong field regimes of relativistic stars could discriminate
between General Relativity and its possible extensions. From compact relativistic
objects data follows, however that general relativity describes gravitation
with very large precision.

There are other classes of dark energy models, such as holographic dark
energy (HDE) \cite{Miao,Qing-Guo,Qing-Guo-2}, which are based on the
holographic principle \cite{Wang,3,4,5} from black hole thermodynamics and
string theory. This principle states that there is a connection between
the infrared cut-off of quantum field theory and the largest distance of
this theory. From the cosmological viewpoint it means that everything in
the universe can be described by some quantities on its boundary. Tsallis
modified the entropy formula for black holes
\cite{Tsallis,Tsallis-2} and created a new class of dark energy model,
called the Tsallis holographic dark energy model (THDE)
\cite{Tavayef,Saridakis,Saridakis-2,Moosavi}.

It should be said that HDE model is significantly different from the other
dark energy models based on scalar-tensor theory or cosmological fluids.
The holographic principle has also been applied to the early inflation
\cite{Horvat,NOS,Paul,Bargach,Timoshkin,Oliveros}. During the early Universe,
the size of the Universe was small, due to which, the holographic energy
density was significant to causing an inflation and it was also found that
this inflation can be matched with the 2018 Planck observations.

THDE models have been studied with different choices of IR cutoffs
\cite{Zadeh} and in different gravity theories
\cite{Ghaffari,Jawad,Nojiri:2019skr}. Authors of
\cite{Nojiri:2017opc} proposed the most generalized holographic dark energy
model with infrared cutoff in form of combination of the Hubble parameter,
particle and events horizons, vacuum energy, the age of universe etc. For
the corresponding choice of the parameters this model is equivalent to
modified gravity or gravity with cosmological fluid. Due to this correspondence,
authors proposed realistic models with inflation or late-time acceleration
in terms of covariant holographic dark energy. Also in models with generalized
HDE it is possible to unify the early inflation with the late cosmological
acceleration. As showed in \cite{Nojiri:2021iko} many HDE models (THDE,
Renyi HDE, and Sharma-- Mittal HDE) are equivalent to the generalized HDE.
It would also be prudent to mention considered models of THDEs on the brane
\cite{AA} and with matter-dark energy interaction \cite{AA2}.

The stability of holographic dark energy is a crucial issue for its viability.
In \cite{Myung} it was found that classic holographic dark energy had a
negative sound speed square, which implied instability. We address to this
question in our paper for generalized Tsallis model, studying the dark
energy perturbations evolution and considering these perturbations from
another viewpoint not as perturbations for cosmological fluid filled universe.

The paper is organized as follows. In the next section we briefly consider
model of holographic dark energy and generalization of this model based
on Tsallis proposition for entropy-area relation. Then we investigate the
evolution of possible perturbations in the model with event horizon as
cut-off. The key moment is that the analysis of perturbations in a case
of holographic dark energy requires another approach than in a case of
normal of cosmological fluid. Because holographic dark energy is a global
quantum phenomenon, the perturbation of the future event horizon should
be calculated as in a case of ordinary holographic dark energy
\cite{Miao}. These calculations are given for various values of non-additivity
parameter for Tsallis dark energy. We firstly consider evolution of metric
perturbations due to perturbations of event horizon neglecting matter perturbations.
But our analysis shows that matter perturbations are very important and
we take into account matter perturbations using approach derived in
\cite{Agos,Silva}. In Section~\ref{sec4} we include in our consideration a possible
realistic interaction between holographic and matter components and investigate
features of perturbations evolution in this case. Finally, we consider
another model of holographic dark energy with Hubble parameter as infrared
cut-off and study perturbations in this case. In conclusion we end this
paper with some discussion of obtained results.

\section{Basic equations}
\label{sec2}

The original representation for holographic dark energy (HDE) follows from
well-known Bekenstein bound for entropy. For entropy and energy density
within volume with characteristical length $L_{0}$ we have the following
inequality
\begin{equation*}
\rho _{de}L_{0}^{4}\leq S,
\end{equation*}
and for entropy the condition $S\sim L_{0}^{2}$ is imposed. Tsallis and
Cirto proposed modified entropy-area relation with account of possible
quantum corrections namely
\begin{equation*}
S = \delta S^{\gamma }= (4\pi )^{\gamma}\delta L_{0}^{2\gamma},
\end{equation*}
where parameter of non-additivity $\gamma $ can differ from $1$. Authors
of \cite{Cohen} founded relation between the entropy, infrared cut-off
($L_{0}$), and the ultraviolet cut-off ($\Lambda $):
\begin{equation*}
L_{0}\Lambda \leq S^{1/4}
\end{equation*}
Tsallis-Cirto relation for entropy gives for ultraviolet cut-off the following
\begin{equation*}
\Lambda ^{4} \leq \delta (4\pi )^{\gamma} L_{0}^{2\gamma - 4}.
\end{equation*}
Based on the HDE hypothesis, $\Lambda ^{4}$ is taken as the density of
dark energy $\rho _{de}$ and therefore
%
\begin{equation}
\rho _{de}=\frac{3C^{2}}{L_{0}^{4-2\gamma }},
\label{eq1}
\end{equation}
where $C^{2} = \delta (4\pi )^{\gamma}/3$ is an unknown parameter with
dimension $M^{2\gamma}$. For scale $L_{0}$ the different choices are proposed.
A simple variant is the Hubble horizon, $H^{-1}$. In \cite{GO},
\cite{GO-2} this model was modified as
\begin{equation*}
L_{0}^{-2} = \alpha H^{2} + \beta \dot{H}.
\end{equation*}
Another possibilities include the particle
\begin{equation*}
L_{0} = a \int _{0}^{t}\frac{d{t}'}{a},
\end{equation*}
and future event horizon:
\begin{equation*}
L_{0} = a \int _{t}^{\infty}\frac{d{t}'}{a}.
\end{equation*}
The infra-red cut-off set by the future event horizon is physically natural
and agrees well with observational data for cosmological acceleration.
We consider mainly this choice.

For universe with Friedmann-Lemetre-Robertson-Walker metric for flat space
%
\begin{equation}
ds^{2} = dt^{2} - a^{2}(t) (dr^{2} + r^{2} d\Omega ^{2}) ,
\label{eq2}
\end{equation}
and filled of HDE and matter Friedmann equation take a form
%
\begin{equation}
\frac{\dot{a}^{2}}{a^{2}} = \frac{1}{3}\left (
\frac{3C^{2}}{L_{0}^{4-2\gamma}} + \rho _{m}\right ).
\label{eq3}
\end{equation}
We also add equation for matter energy density $\rho _{m}$ which follows
from Einstein equations:
%
\begin{equation}
\dot{\rho}_{m} + 3\frac{\dot{a}}{a}\left (\rho _{m} + p_{m}\right ) = 0.
\label{eq4}
\end{equation}
Solving these equations and equation for $L_{0}$ we can obtain evolution
of scale factor $a$ with time.

The evolution of possible fluctuations of HDE density merits further investigation.
For $\gamma =1$ corresponding calculations have being made in
\cite{Myung}. At first glance, for HDE as fluid component square of sound
speed is negative and therefore perturbations of HDE are unstable. But
one needs to account that HDE is given by the holographic vacuum energy
whose perturbation should be considered globally. In the next section we
perform calculations for perturbations of HDE using approach presented
in \cite{Miao-2} for $\gamma =1$.

\section{Evolution of perturbations in model of Tsallis HDE with event horizon as cut-off}
\label{sec3}

For simplicity, we consider only scalar perturbations of metric. In Newtonian
gauge we have the following expression for perturbed metric:
%
\begin{equation}
ds^{2} = -\left [ 1 + 2 \Phi (r,t) \right ]dt^{2} + a^{2}(t)\left [ 1 -
2 \Phi (r,t) \right ]dr^{2},
\label{eq5}
\end{equation}
where function $\Phi $ depends not only on time but on comoving radial
coordinate. The physical distance $L(0,t)$ for horizon from observer locating
at $r=0$ can be found from integral
%
\begin{equation}
L(0,t) = \int _{0}^{l(0,t)} a(t)\left [ 1 - \Phi (r,t) \right ] dr,
\label{eq6}
\end{equation}
where $l(0,t)$ is the coordinate distance to the future event horizon
%
\begin{equation}
l(0,t) \equiv l_{0} + \delta l, \: \: \: \delta l = \int _{t}^{\infty}
\frac{2 \Phi (l_{0}(t'),t' )}{a(t')}dt'.
\label{eq7}
\end{equation}
The value $l_{0}$ means comoving unperturbed distance to event horizon,
\begin{equation*}
l_{0} = \int _{t}^{\infty} \frac{dt'}{a(t')}dt'.
\end{equation*}
Thus the fluctuation of the future event horizon can be written as
%
\begin{equation}
\delta L \equiv L(0,t) - L_{0} =a(t)\left [ \int _{t}^{\infty}
\frac{2 \Phi (l_{0}(t'),t' )}{a(t')}dt' - \int _{0}^{l_{0}} \Phi (r,t)dr
\right ]
\label{eq8}
\end{equation}
and the THDE energy density has corresponding fluctuation
%
\begin{equation}
\delta \rho _{de} = ({2\gamma -4})\rho _{de} \frac{\delta L}{L_{0}}
\label{eq9}
\end{equation}
Inserting this equation into the 00-component of the perturbated Einstein
equation, one obtains equation for function $\Phi (r,t)$
%
\begin{equation}
\frac{\Delta ^{2}}{a^{2}}\Phi - 3H\dot{\Phi} - 3H^{2}\Phi =
\frac{1}{2}(\delta \rho _{de} + \delta \rho _{m})
\label{eq10}
\end{equation}

Firstly we consider the evolution of the universe from current epoch when
perturbations of matter are negligible. We only take into account possible
fluctuations of dark energy and investigate its evolution with time. In
this case we can obtain one independent equation for $\Phi (r,t)$. Perturbations
of dark energy density are defined by $\Phi (r,t)$.

To solve equation (\ref{eq10}) in a case when $\delta \rho _{m} = 0$ we expand
$\Phi $ using its eigenfunction. We suggest
%
\begin{equation}
\Phi (r,t) = \sum _{k}\Phi _{k}(t)\frac{\sin (kr)}{r}
\label{eq11}
\end{equation}
%

\begin{figure*}
\includegraphics[scale=0.3]{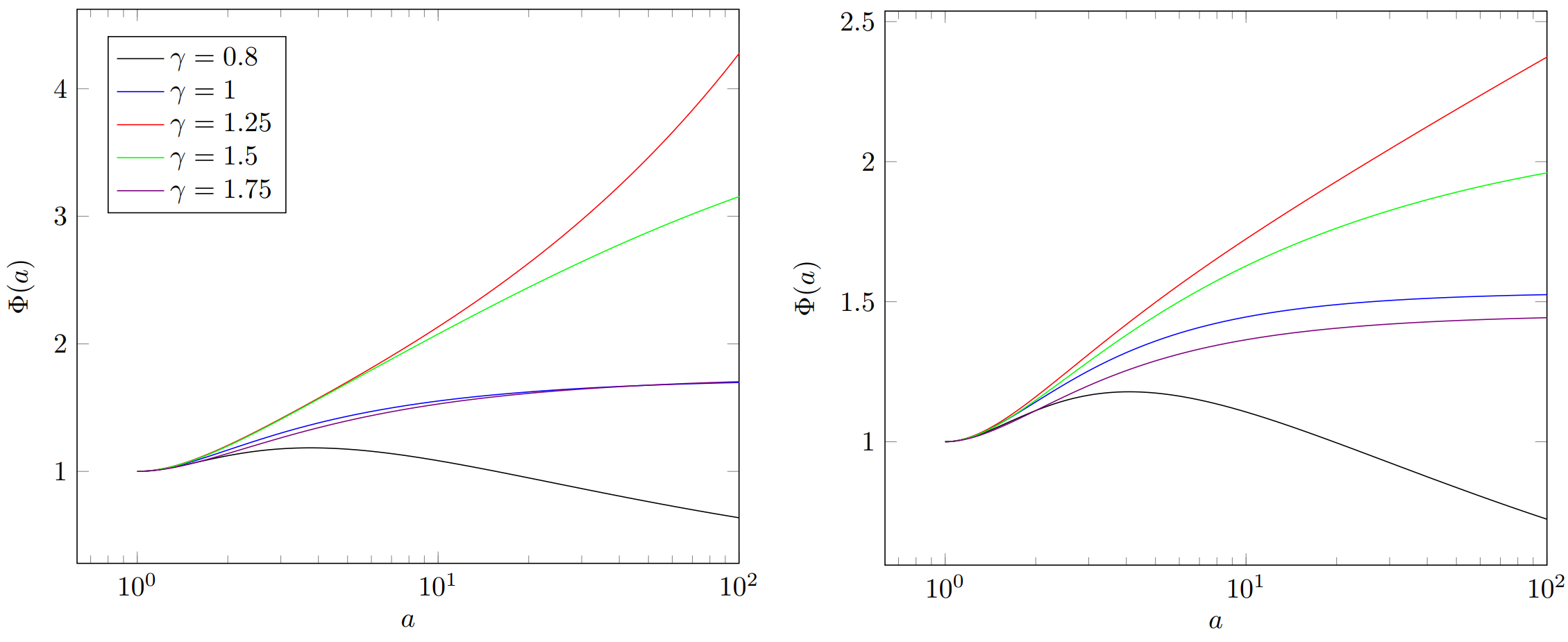}
\caption{Evolution of perturbations for metric in future as function of scale
factor for $\Omega _{de0}=0.72$, $C=0.8$ (left panel), $C = 1$ (right panel) for
various values of parameter $\gamma $.}
\label{fig1_v1}
\end{figure*}

where we have dropped the $\cos (kr)/r$ terms, which lead to a singularity
at $r = 0$. One way to deal with this equation is to take derivative with
respect to $t$. This integral equation becomes a differential equation
%
\begin{equation}
\label{eqPhi}
\begin{aligned}
\ddot{\Phi}_{k}
&+ \frac{1}{3H}
\Bigg( \frac{k^{2}}{a^{2}} + 3\dot{H}
+3H^{2}+ 3(4 - 2\gamma )H^{2} - \frac{3H(5-2\gamma )}{L_{0}}
\\&
+\frac{4-2\gamma}{2} \frac{a\rho _{de}}{L_{0}} \int _{0}^{l_{h0}(t)}
\frac{\sin (kr)}{kr}dr \Bigg) \dot{\Phi}_{k}
+\frac{1}{3H} \Bigg(6\dot{H}H
+ 3(4 - 2\gamma )H^{3}
\\&
+ (2 - 2\gamma )H
\frac{k^{2}}{a^{2}}
- \frac{3H^{2}(5-2\gamma )}{L_{0}} - (5-2\gamma )
\frac{k^{2}}{a^{2}L_{0}}
\\&
+ \frac{4-2\gamma}{2}
\frac{\rho _{de}}{kL_{0}} \frac{\sin (kl_{0}(t))}{l_{0}(t)} \Bigg)
\Phi _{k} = 0.
\end{aligned}
\end{equation}
This equation for function $\Phi _{k}$ should be solved with equation for
$L_{0}$ for the case when $L_{0}$ is the event horizon:
\begin{equation*}
\dot{l}_{0} = -\frac{1}{a},\quad l_{0} = \frac{L_{0}}{a},
\end{equation*}
and Friedmann equations
\begin{equation*}
\dot{a} = Ha
\end{equation*}
\begin{equation*}
\dot{H} = -\frac{1}{2}\left ( \rho _{m} + \frac{1}{3} (4-2\gamma )
\rho _{de} \left ( 1 - \frac{1}{L_{0} H} \right )\right )
\end{equation*}
Integration of equation for matter density gives a well-known relation
\begin{equation*}
\rho _{m} = \frac{D}{a^{3}},
\end{equation*}
where $D$ is integration constant which can be defined from initial conditions.
For that purpose, it is convenient to choose initial values of overall
density and scale factor
\begin{equation*}
a(0)=1, \quad \rho = \rho _{de}+\rho _{m} = 1.
\end{equation*}
This choice gives Hubble parameter at $t=0$ value $1/\sqrt{3}$. Equation
for $\Phi _{k}$ is invariant under scaling transformation, therefore we
can assume that $\Phi _{k}(0) = 1$. If $\Phi _{k}>1$ for $t>0$ perturbations
grow in future. Currently, $\rho _{de}\approx 0.7$ and
$\rho _{m}\approx 0.3$. We estimate $D=0.28$ and find the corresponding
initial value of $L_{0}$.

Firstly, we consider perturbations in the case of Tsallis HDE without interaction
between matter and dark energy. For brevity we omit index $k$ from
$\Phi _{k}$ and consider evolution of mode with $k=0.01$. Our calculations
show that dependence of $\Phi _{k}(t)$ on value of $k$ is very negligible.
We see the following features (see Fig.~\ref{fig1_v1}). For $C=1$ perturbations drop
down for $\gamma <1$, but for $\gamma >1$ after some growth function
$\Phi $ asymptotically tends to the constant. For $C>1$ the picture is
the same. If $C<1$ for some $\gamma $ perturbations increase. Again, as
in for $C\geq 1$, for $\gamma <1$ perturbations vanish.

\begin{figure*}[t]
\includegraphics[scale=0.3]{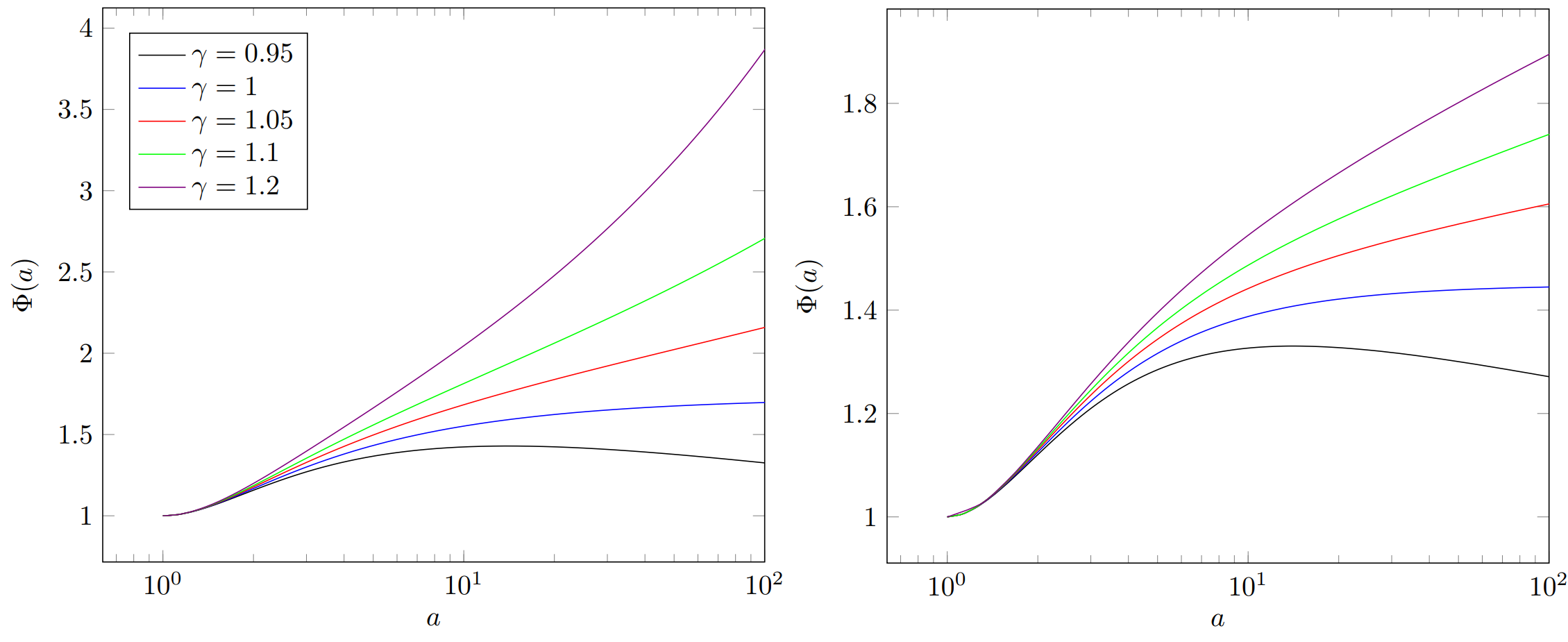}
\caption{Evolution of perturbations in future with as function of scale factor
for $\Omega _{de0} = 0.72$, $C = 0.8$ (left panel), $C=1.2$ (right panel) for
various values of parameter $\gamma $.}
\label{fig2_v1}
\vspace{10pt}
\end{figure*}

\begin{figure*}
\includegraphics[scale=0.3]{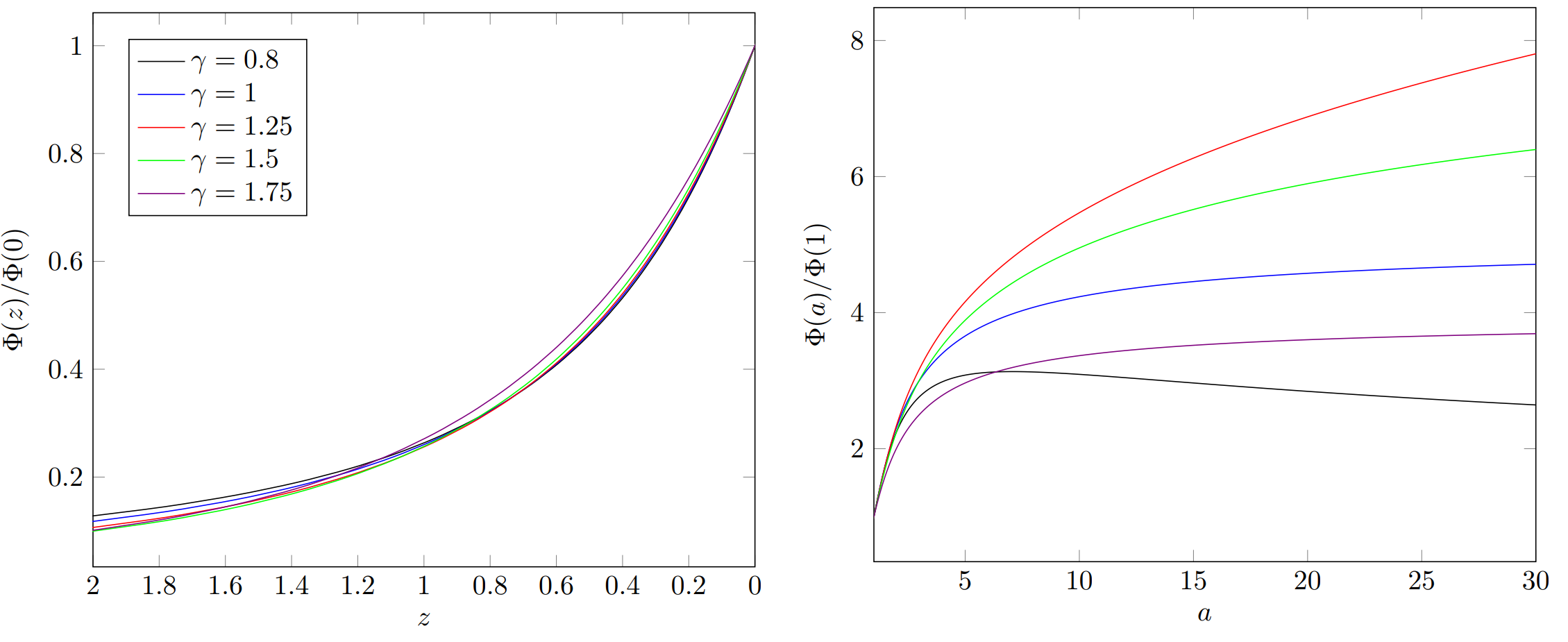}
\includegraphics[scale=0.3]{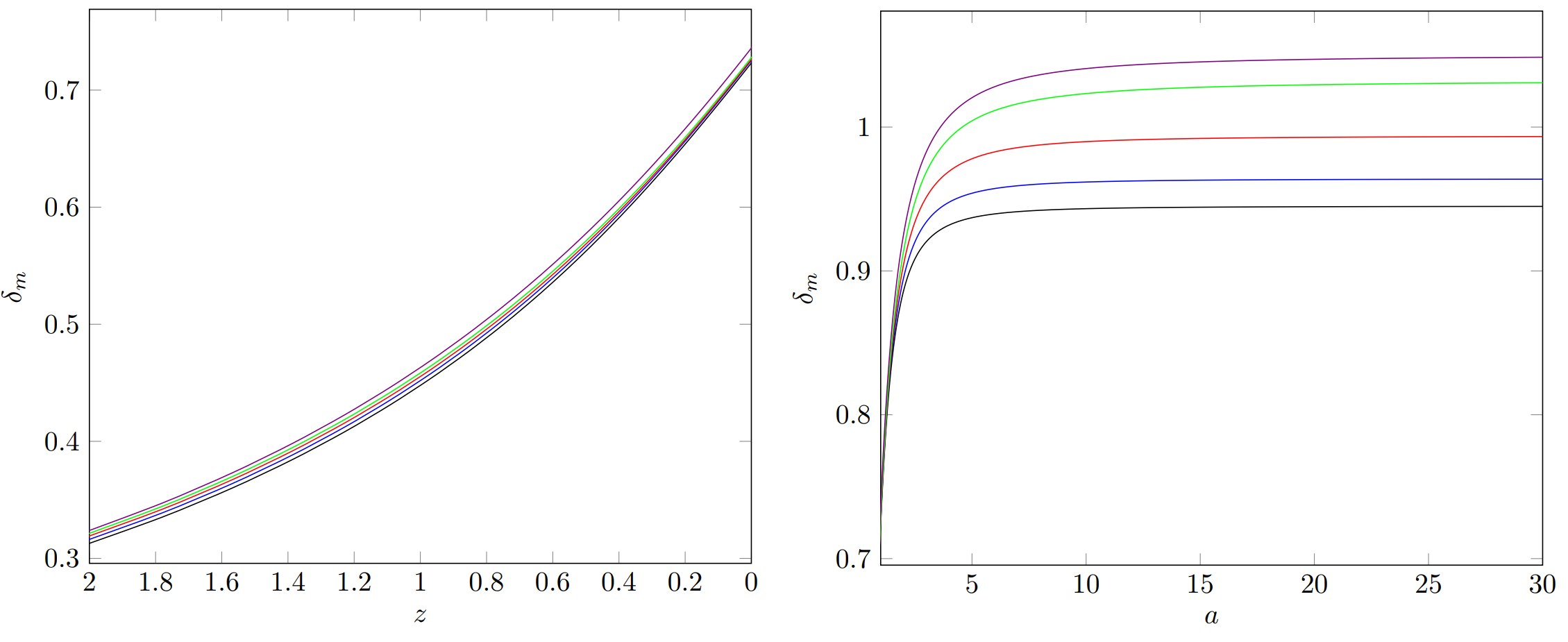}
\caption{Evolution of metric perturbations in past (upper left panel) as function
of redshift $z$ in interval $0<z<2$ and in future as function of scale factor
(upper right panel). On down panel the corresponding evolution of matter
perturbations is given for past and in future. For current value of $\Omega
_{de}$ we take 0.72. Parameter $C = 0.8$.}
\vspace{15pt}
\label{fig3}
\end{figure*}

\begin{figure*}[t]
\includegraphics[scale=0.3]{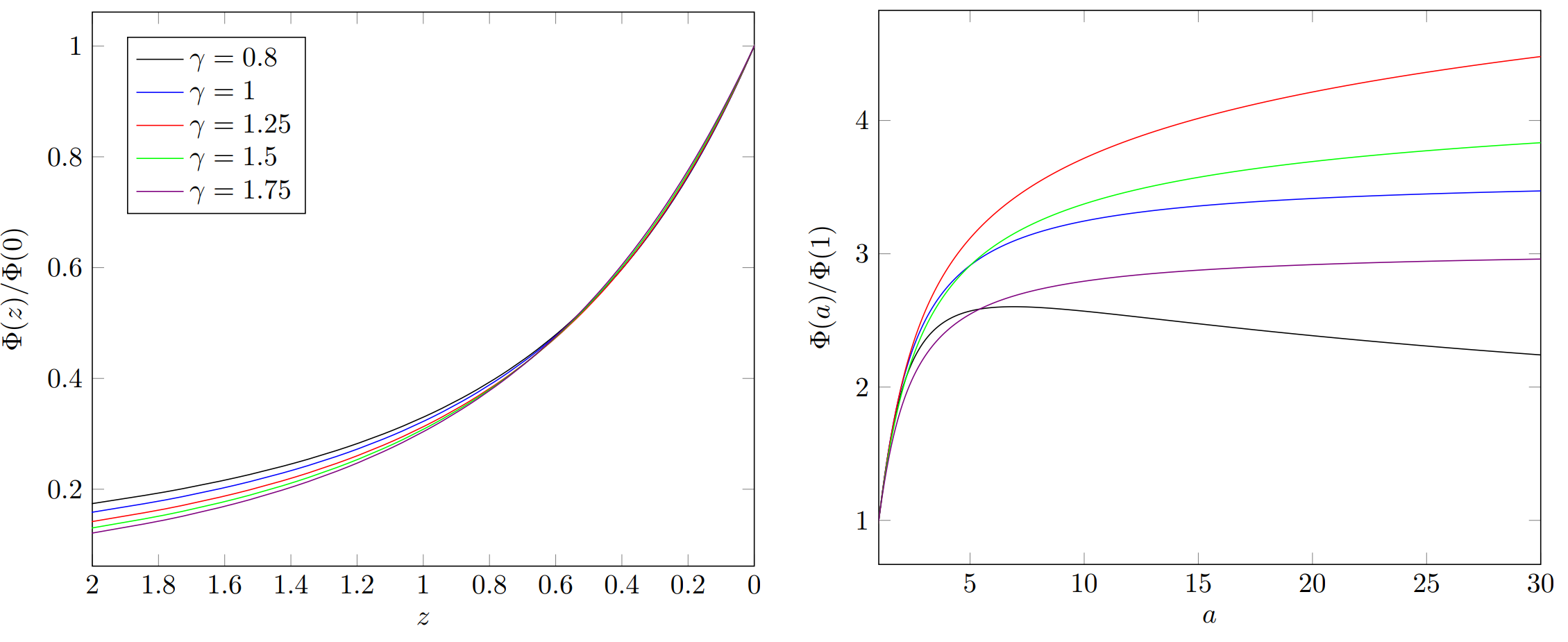}
\includegraphics[scale=0.3]{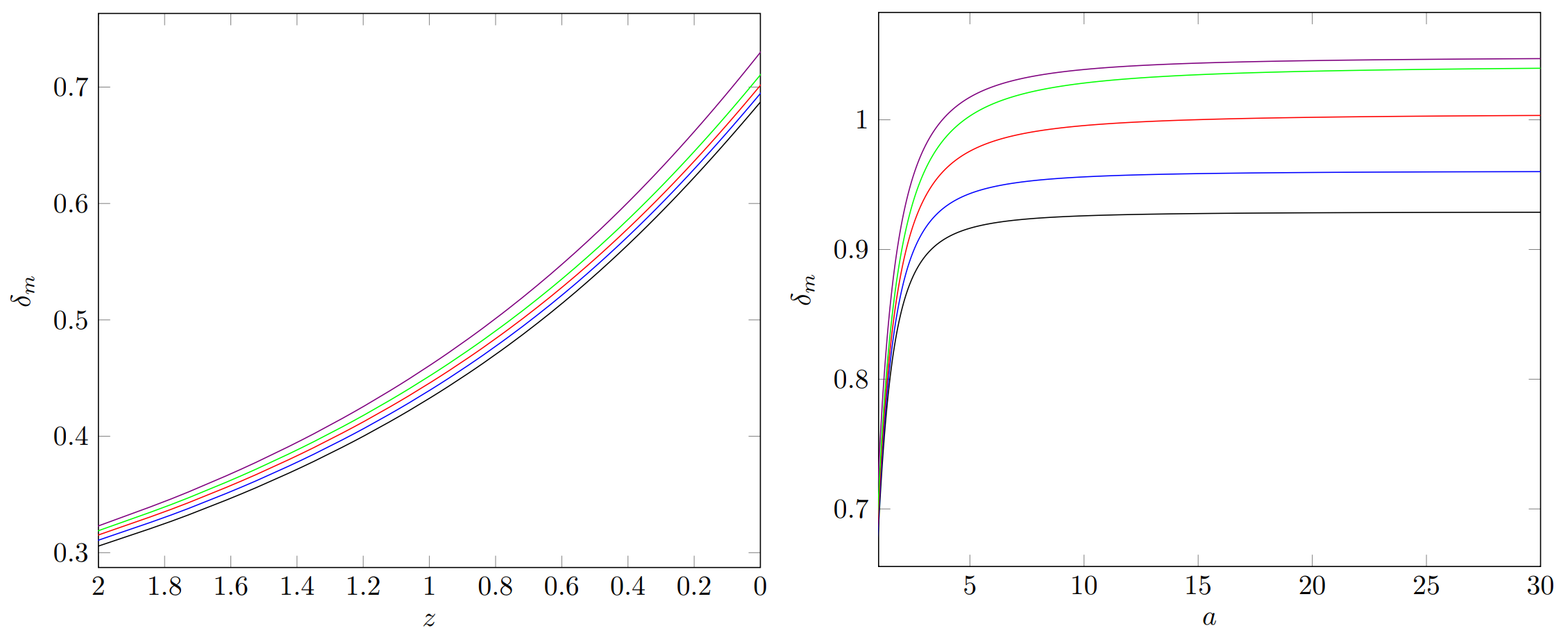}
\caption{The same as on Fig.~\ref{fig3} but for $C=1$.}
\label{fig4}
\end{figure*}

\begin{figure*}[t]
\includegraphics[scale=0.3]{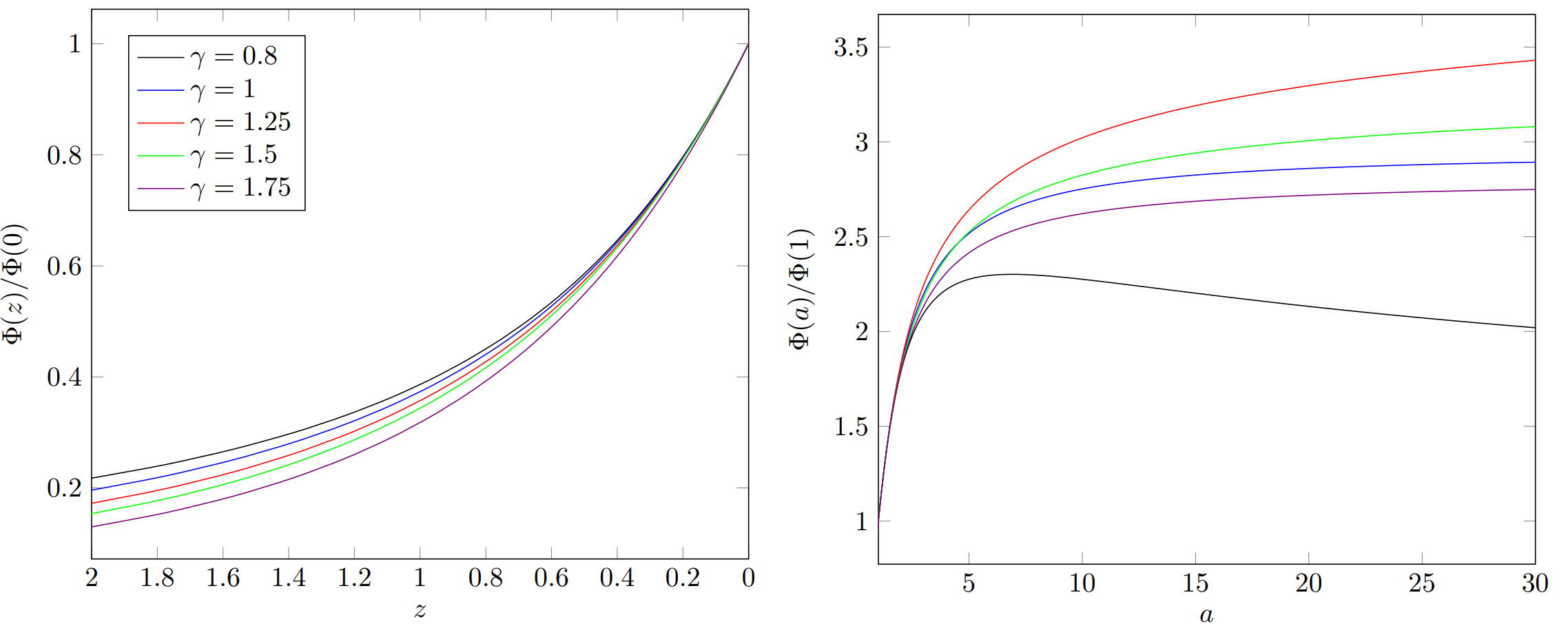}
\includegraphics[scale=0.3]{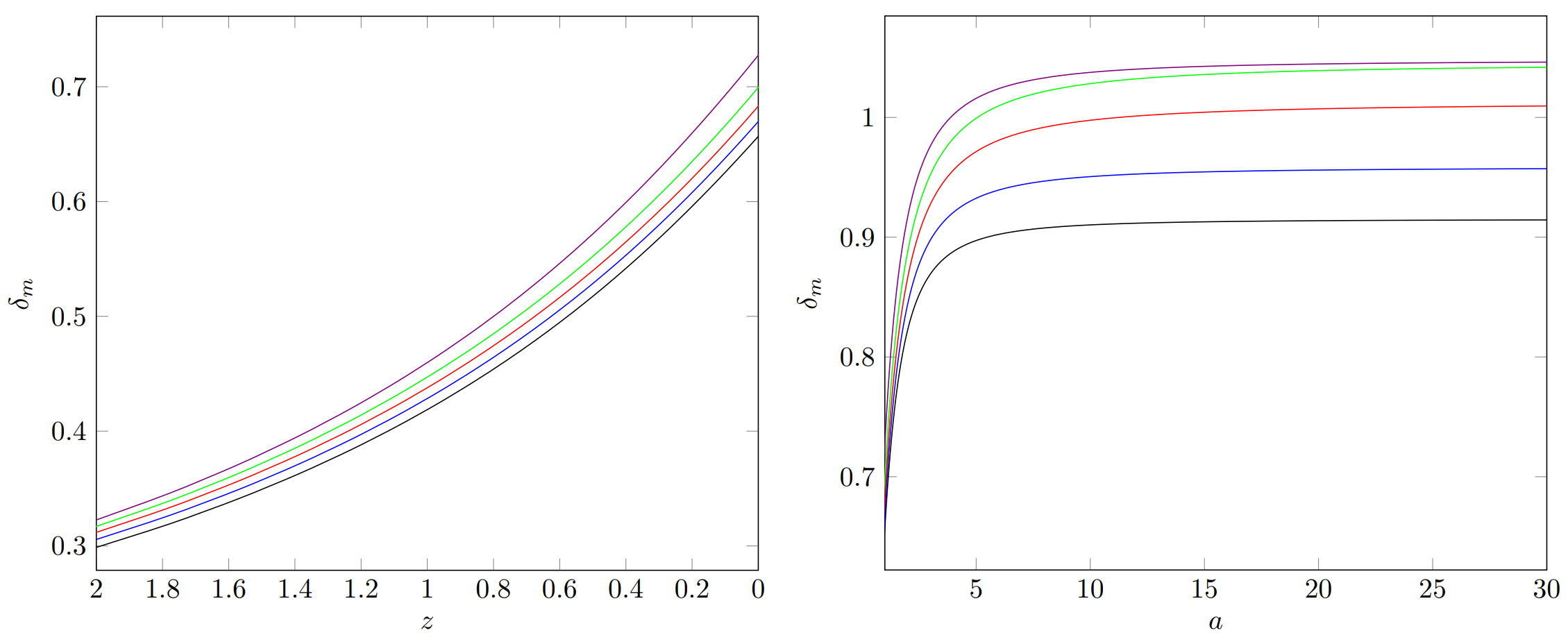}
\caption{The same as on Fig.~\ref{fig3} but for $C=1.2$.}
\label{fig5}
\end{figure*}

There is no unambiguous dependence between value of $\gamma $ and growth
of perturbations for $C<1$ and $C\geq 1$. From Fig.~\ref{fig1_v1} it can be seen that
for $\gamma $ greater than some value the growth rate of perturbations
decreases. Numerical calculations allow to conclude that for
$1<\gamma < \approx 1.3$ perturbations grow infinitely for $C=0.8$ (see
Fig.~\ref{fig2_v1}). The critical value of $\gamma $ for $C<1$ depends from $C$.

\section{Contribution of matter perturbation}
\label{sec4}

Possible evolution of scalar perturbations deserves to be studied not only
at its current state, but in its distant past. Therefore we turn to the
early times when $\Omega _{de}<<1$. In this case one needs to account perturbations
of matter in r.h.s. of equation (\ref{eq10}). The following additional term in
equation for $\Phi $ is
%
\begin{equation}
-\frac{1}{2}\frac{d}{dt}\left ( \frac{\delta \rho _{m}}{\rho _{de}a}L_{0}
\right )\frac{\rho _{de}a}{L_{0}} = -\frac{1}{2}\delta \dot{\rho}_{m}-
\frac{1}{2}\delta \rho _{m}\left ( (4-2\gamma )H -
\frac{5-2\gamma}{L_{0}} \right )
\label{eq13}
\end{equation}
where
%
\begin{equation}
\delta \dot{\rho}_{m} = \rho _{m}\left ( \dot{\delta}_{m} - 3H\delta _{m}
\right ), \quad \delta _{m} = \delta \rho _{m} / \rho _{m}.
\label{eq14}
\end{equation}
Assuming that for relative fluctuation of matter the following representation
is valid
\begin{equation*}
\delta _{m} (r,t) = \sum _{k}\delta _{mk}(t)\frac{\sin (kr)}{r}
\end{equation*}
we obtain the equation for metric perturbations $\Phi _{k}$
%
\begin{equation}
\label{eqPhi_2}
\begin{aligned}
\ddot{\Phi}_{k} &+ \frac{1}{3H} \Bigg( \frac{k^{2}}{a^{2}} + 3\dot{H} +
3H^{2} + 3(4 - 2\gamma )H^{2} - \frac{3H(5-2\gamma )}{L_{0}}
\\&
+
\frac{4-2\gamma}{2} \frac{a\rho _{de}}{L_{0}} \int _{0}^{l_{h0}(t)}
\frac{\sin (kr)}{kr}dr \Bigg) \dot{\Phi}_{k}+
\frac{1}{3H} \Bigg(6\dot{H}H
\\&
+ 3(4 - 2\gamma )H^{3} + (2 - 2\gamma )H
\frac{k^{2}}{a^{2}}
- \frac{3H^{2}(5-2\gamma )}{L_{0}}
\\&
- (5-2\gamma )
\frac{k^{2}}{a^{2}L_{0}} + \frac{4-2\gamma}{2}
\frac{\rho _{de}}{kL_{0}} \frac{\sin (kl_{0}(t))}{l_{0}(t)} \Bigg)
\Phi _{k}
\\=
&%
-\frac{1}{6H}\dot{\delta}_{mk} -\frac{1}{6H}\delta _{mk} \rho _{m}
\left ( (1-2\gamma )H - \frac{5-2\gamma}{L_{0}} \right ).
\end{aligned}
\end{equation}
For relative fluctuations of matter density we have for sub-horizon scales
($k^{2}>>a^{2} H^{2}$) the following equation for $\delta _{m}$ as function
scale factor:
%
\begin{equation}
{\delta}''_{m} + A_{m}\delta '_{m}= S_{m},
\label{eq16}
\end{equation}
where
%
\begin{equation}
A_{m} = \frac{3}{2a}\left ( 1 - \delta _{de}\frac{\rho _{de}}{3H^{2}}
\right ) = \frac{3}{2a}\left ( 1 - \frac{p_{de}}{3H^{2}} \right )
\label{eq17}
\end{equation}
and
%
\begin{equation}
S_{m} = \frac{3}{2a^{2}} \left ( \Omega _{m}\delta _{m} + \left ( 1 + 3
\frac{dp_{de}}{d\rho _{de}} \right )\Omega _{de}\delta _{de}\right ).
\label{eq18}
\end{equation}
The prime denotes the derivative on scale factor. We neglect perturbations
of dark energy for matter perturbations assuming that
$\delta _{de}=0$ in expressions for $A_{m}$ and $S_{m}$ and therefore
\begin{equation*}
A_{m} = \frac{3}{2a}, \quad S_{m} = \frac{3}{2a^{2}}\Omega _{m}
\delta _{m}.
\end{equation*}

We investigate evolution of matter density and corresponding metric fluctuations
from early times when $a_{in}=0.01$. For $\delta _{m}$ we assume that in
this moment $\delta _{m}(a_{in})=0.01$ and
$\delta '_{m}(a_{in}) = 1$). With account of matter perturbations there
is no scale invariance in equation for $\Phi _{k}$. For simplicity we take
as initial conditions for $\Phi _{in}$ and $\Phi '_{in}$ $10^{-4}$ and
$0$ correspondingly. Then we integrate equations from $a=0.01$ to
$a=30$ (distant future when $\Omega _{de} \rightarrow 1$). We found
$\Phi $ at current value of scalar factor ($a=1$) and consider the relation
$\Phi (a)/\Phi (1)$ as function of redshift $z$ in past (in range
$0<z<2$) and as function of scale factor in future.

Results of our calculations for various values of $\gamma $ and
$C=0.8$, $1.0$, $1.2$ are given on Figs.~\ref{fig3}, \ref{fig4}, \ref{fig5}. From our calculations
we see that evolution of matter perturbations doesn't depend significantly
from parameter $C$, but value of non-additivity parameter affects considerably
on asymptotical value of $\delta _{m}$ on large times. For smaller values
of $\gamma $ the limit
$\delta _{ms}=\lim _{t\rightarrow \infty} \delta _{m}$ decreases. We see
also that evolution of metric perturbations significantly changes in comparison
with case when we neglect matter perturbations. There is no significant
difference in principal character of evolution of $\Phi $ in past and future
only some details change. For all values of $C$ metric fluctuations freeze
with time in future but the growth of fluctuations depends strongly from
parameter $\gamma $. Also as in a case of neglecting matter perturbations
there are no simple correlation between $\gamma $ and asymptotical value
of $\Phi (a)$ at large $a$.

\section{Perturbations in a case of the interaction between dark energy and matter}
\label{sec5}

Let's consider the model of Tsallis HDE with interaction between dark energy
and matter. The interaction between the two components can be introduced
by the following. Equations of continuity for components 1 and 2 are
%
\begin{equation}
\dot{\rho}_{1} + 3H (\rho _{1} + p_{1}) = 0, \quad \dot{\rho}_{2} + 3H
(\rho _{2} + p_{2}) = 0,
\label{eq19}
\end{equation}
and are modified as
%
\begin{equation}
\dot{\rho}_{1} + 3H (\rho _{1} + p_{1}) = Q_{12},\quad \dot{\rho}_{2} +
3H (\rho _{2} + p_{2}) = -Q_{12},
\label{eq20}
\end{equation}
where $Q_{12}$ is some function of density of components, time, Hubble
parameter, et cetera, in general case. We assume simple possibility for
interaction between holographic component of dark energy and matter which
are described by the function
\begin{equation*}
Q = H({\alpha} \rho _{m} + \beta \rho _{de}).
\end{equation*}
Here $\alpha $, $\beta $, $\lambda $ are some dimensionless constants.
For Tsallis HDE we can find pressure using continuity equation with interaction
term and obtain the following equation for $\dot{H}$:
%
\begin{equation}
\dot{H} = -\frac{1}{2}\left ( \rho _{m} + \frac{Q}{3H} + \frac{1}{3} (4-2
\gamma )\rho _{de} \left ( 1 - \frac{1}{LH} \right )\right ).
\label{eq21}
\end{equation}
The pressure of matter is zero and for density of matter one needs to solve
the equation:
%
\begin{equation}
\dot{\rho}_{m} + 3H\rho _{m} + Q = 0.
\label{eq22}
\end{equation}

\begin{figure*}
\includegraphics[scale=0.3]{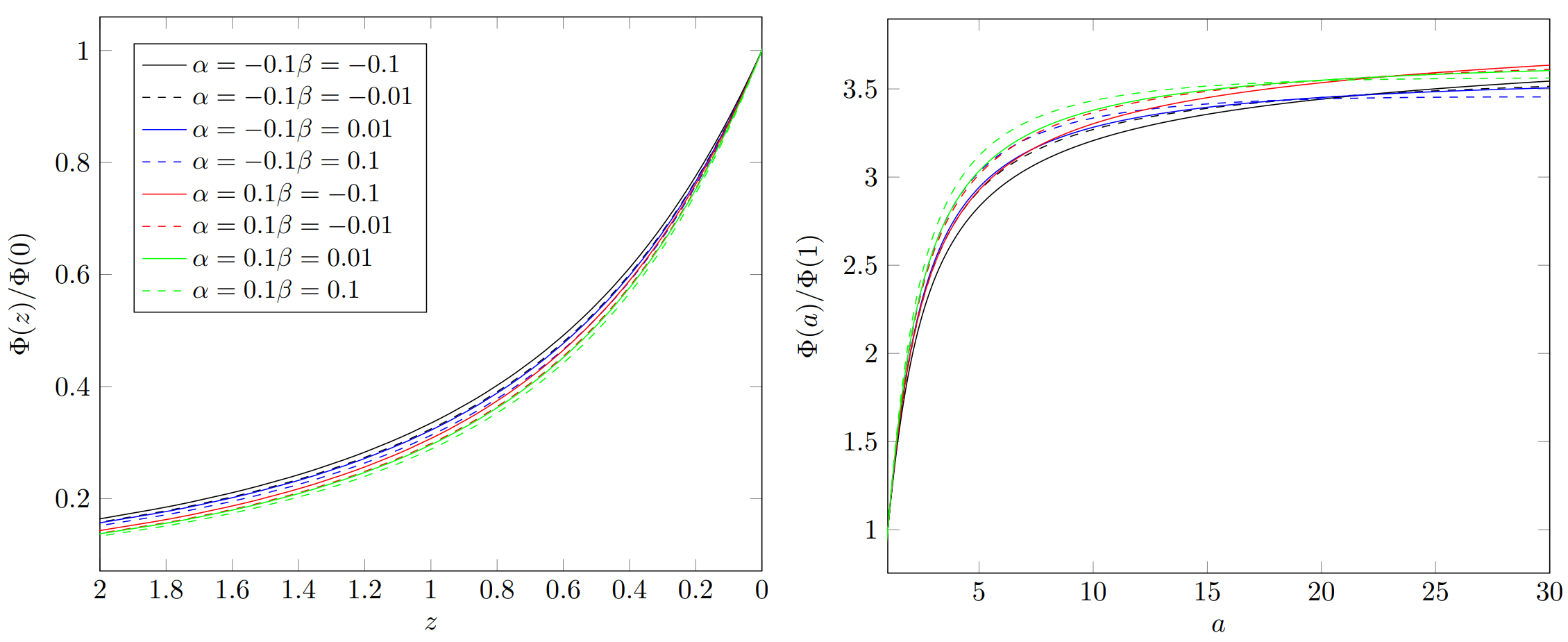}
\includegraphics[scale=0.3]{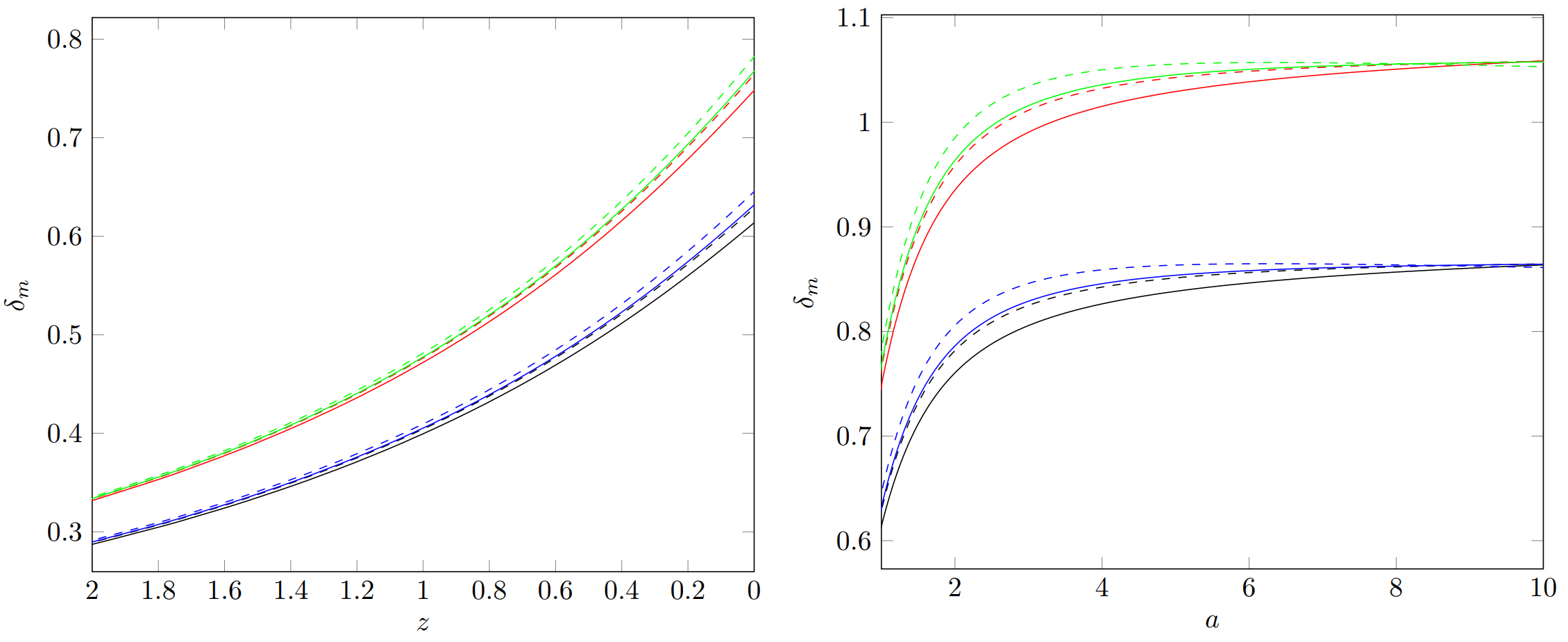}
\caption{Evolution of metric perturbations in past (upper left panel) as
function of redshift and in future as function of scale factor (upper right
panel) in a case of interaction between matter and HDE with $Q = H(\alpha \rho
_{m}+\beta \rho _{de})$ for various values $\alpha $ and $\beta $. The other
parameters are $C = 1,\gamma = 1$. For current value of $\Omega _{de}$ we take
$0.72$. On down panel the corresponding evolution of matter perturbations is
depicted. This evolution is sensitive to sign of parameter $\alpha $.}
\label{fig7}
\end{figure*}

\begin{figure*}[t]
\includegraphics[scale=0.3]{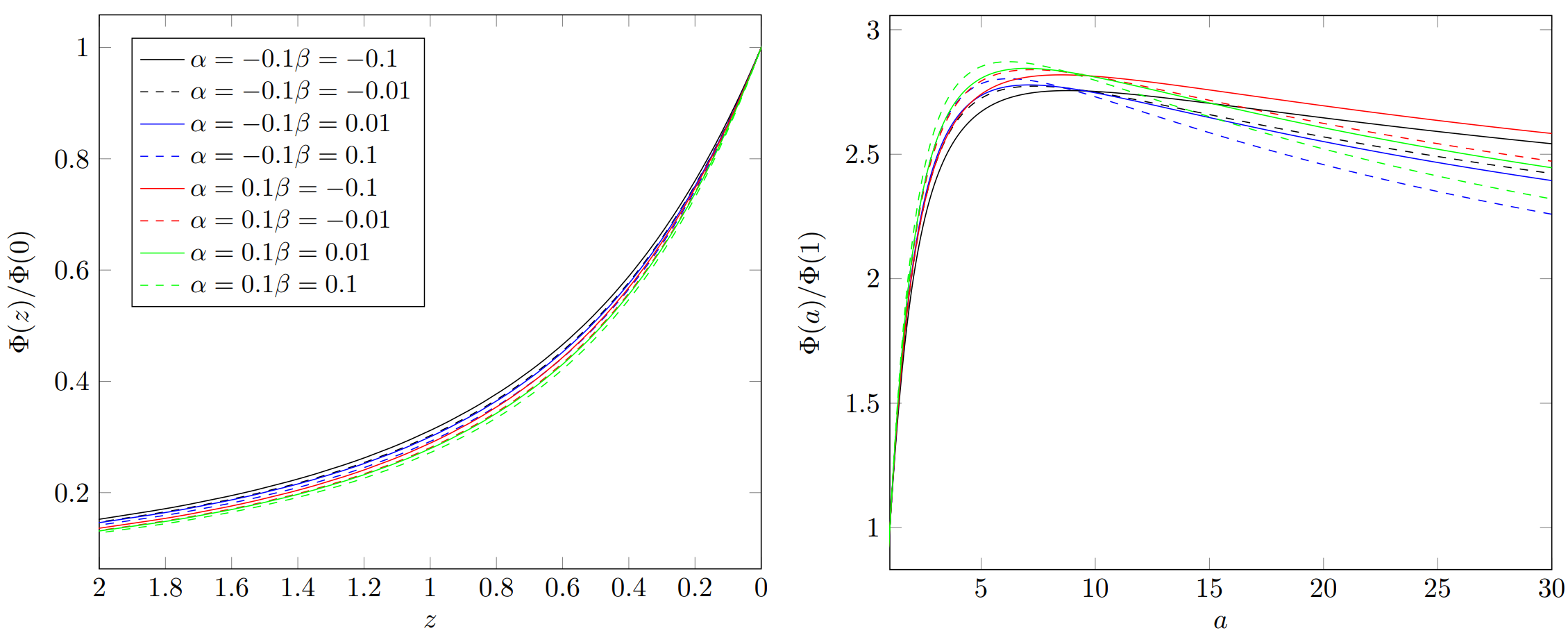}
\includegraphics[scale=0.3]{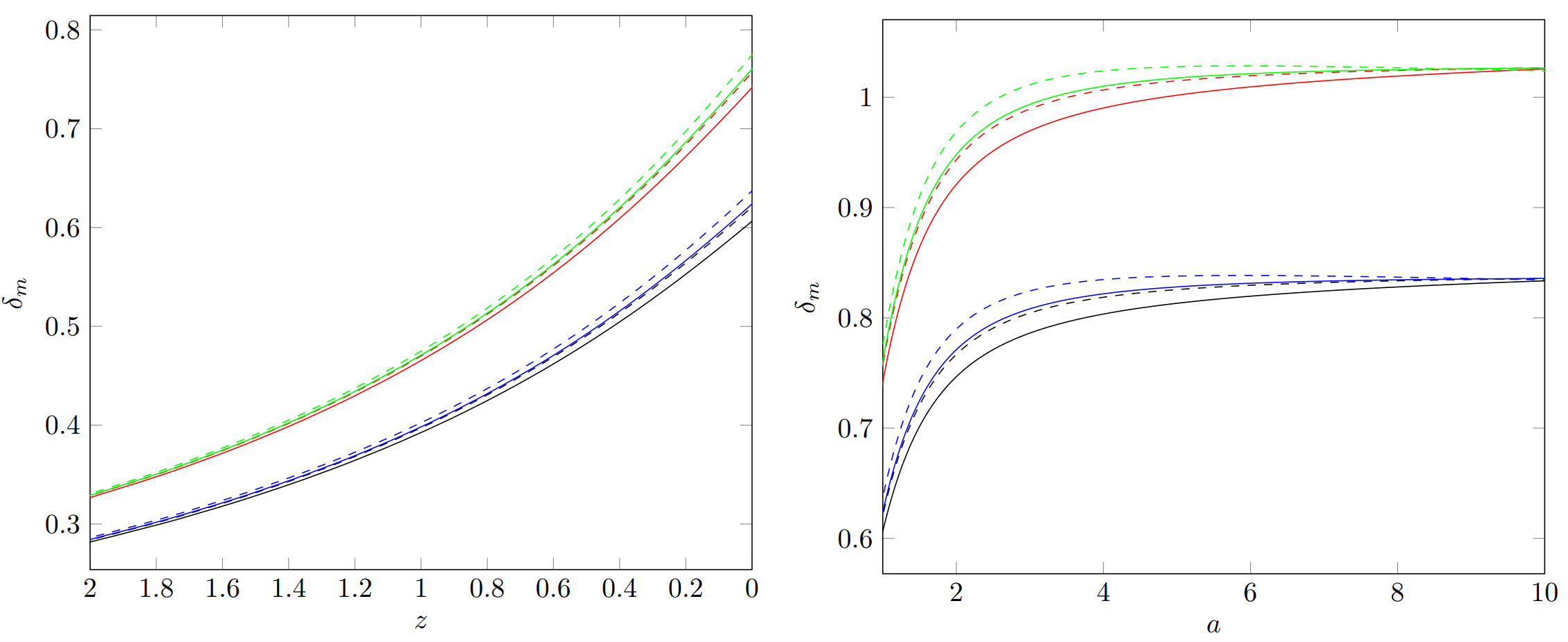}
\caption{The same as on Fig.~\ref{fig7}, but for $C=1,\gamma =0.8$.}
\label{fig8}
\end{figure*}

\begin{figure*}[t]
\includegraphics[scale=0.3]{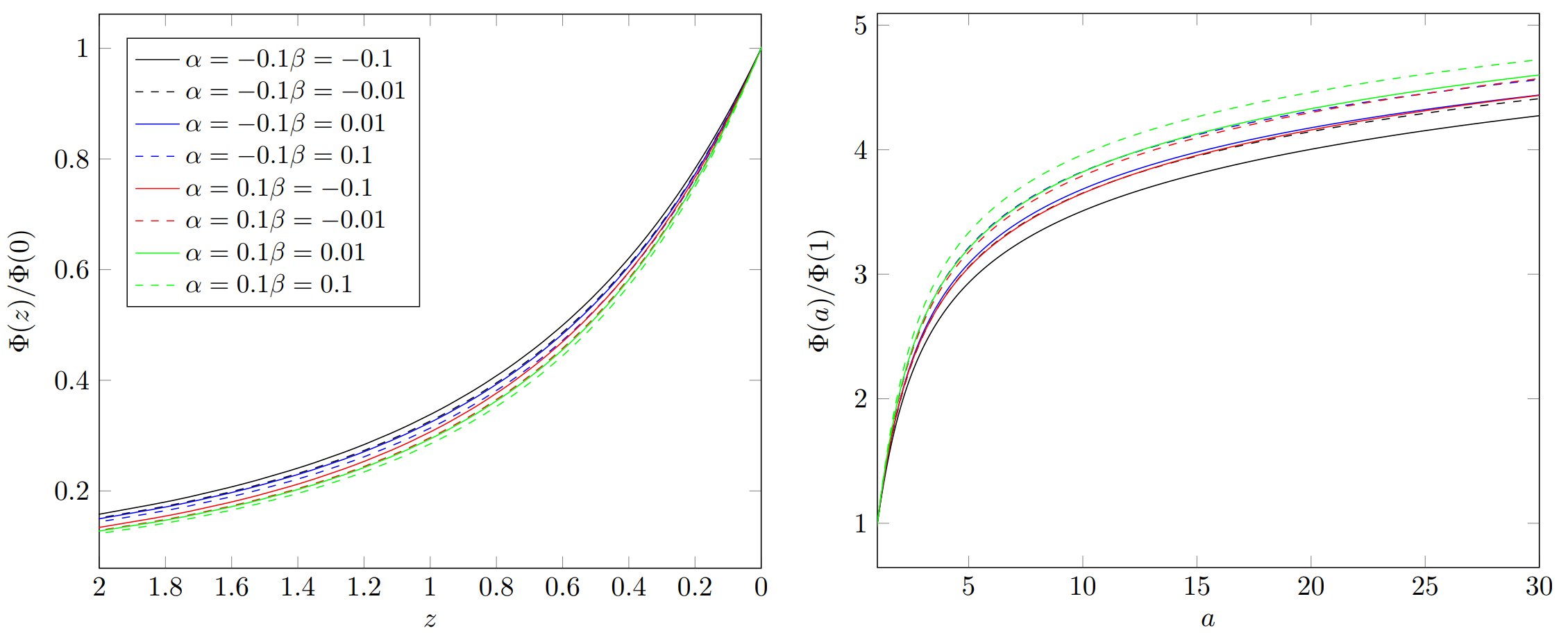}
\includegraphics[scale=0.3]{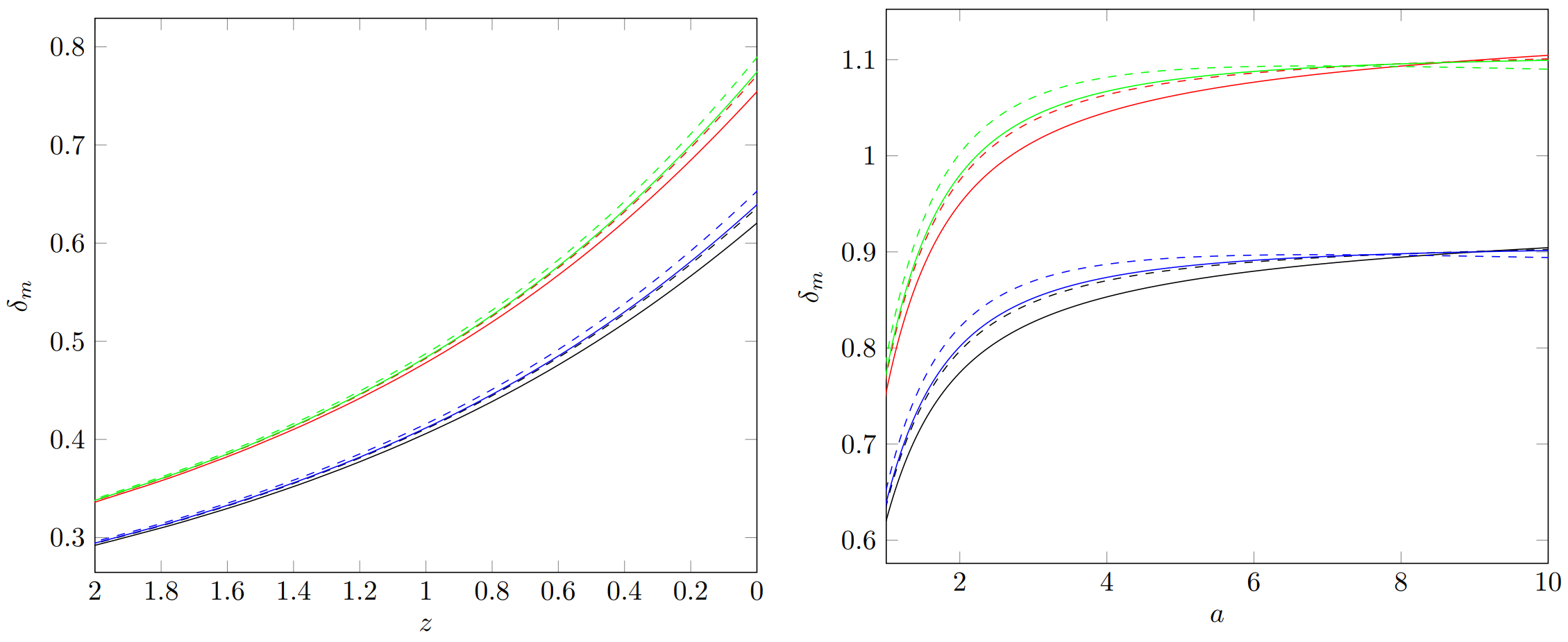}
\caption{The same as on Fig.~\ref{fig7}, but for $C=1,\gamma =1.2$.}
\label{fig9_v1}
\end{figure*}

Firstly we consider the evolution of metric perturbations without contribution
of matter perturbations using Eq. (\ref{eqPhi}) for $\Phi _{k}(t)$. The results for
various values of $\alpha $ and $\beta $ are presented on Fig.~\ref{fig7} for
$\gamma =1.0$ and $C=1$. For various values of $\gamma $ close to
$1$ perturbations decay with time (see Figs.~\ref{fig8} and \ref{fig9_v1}). For some $\alpha $ and $\beta $ solutions
with interesting behavior are possible: perturbations initially increase,
then decrease and after some increasing they decay finally.

For $C<1$ and $C>1$ function for some $\alpha $ and $\beta $ solutions
with interesting behavior are possible: perturbations initially increase,
then decrease and after some increasing they decay finally.

If we include matter perturbations in our calculations situation dramatically
changes. For metric perturbations we have only numerical but not qualitative
difference between cases with different values of $C$ and $\gamma$. The
function $\Phi $ asymptotically tends to constant in all cases. The evolution
of matter perturbations depends strongly from sign of parameter
$\alpha $. For positive $\alpha $ growth of matter perturbations is more
than for $\alpha <0$.

\section{Evolution of perturbation with inverse Hubble scale as cut-off}
\label{sec6}

Another interesting choice for cut-off is a Hubble horizon. In this case
for density of dark energy we have
\begin{equation*}
\rho _{de} = 3C^{2} H^{4-2\gamma} = 3C^{2}
\frac{\dot{a}^{4-2\gamma}}{a^{4-2\gamma}}
\end{equation*}
Using relations for $\delta a$ and $\delta \dot{a}$
\begin{equation*}
\delta a = - \Phi (r,t)a, \quad \delta \dot{a} = - \dot{\Phi}(r,t)a -
\Phi (r,t)\dot{a}
\end{equation*}
one can obtain the following equation for perturbations of dark energy
density
%
\begin{equation}
\delta \rho _{de} = (4-2\gamma )\rho _{de}\left (\Phi (r,t) -
\frac{\dot{\Phi}(r,t)}{H}\right ){,}
\label{eq23}
\end{equation}
and corresponding equation for modes $\Phi _{k}$:
\begin{equation*}
\left ( 3H^{2} + (\gamma - 2)\rho _{de}) \right ) \dot{\Phi}_{k} +
\left ( 3H^{3} + \frac{k^{2}}{a^{2}}H \right )\Phi _{k} = 0.
\end{equation*}

There are two ways of cosmological evolution for such model of holographic
dark energy. Firstly, when Hubble parameter approaches to some non-zero
constant value. Secondly, scale factor grows and therefore
\begin{equation*}
H^{2}\approx \rho _{de}/3=C^{2} H^{4-2\gamma},
\end{equation*}
and for $t\rightarrow \infty $ Hubble parameter
\begin{equation*}
H \rightarrow C^{\frac{1}{\gamma -1}}.
\end{equation*}
From equation of perturbations written in form
\begin{equation*}
\dot{\Phi}_{k} = -
\frac{H+k^{2}a^{-2}H^{-1}/3}{1+(\gamma -2)\frac{\rho _{de}}{3H^{2}}}
\Phi _{k}
\end{equation*}
we can conclude that for $\gamma \leq 1$ in the case of HDE-dominated universe
perturbation increase infinitely. For $\gamma >1$ $\Phi _{k}$ exponentially
decays
\begin{equation*}
\Phi _{k} \sim \exp (-\alpha t), \quad \alpha =
\frac{C^{\frac{1}{1-\gamma}}}{\gamma - 1}.
\end{equation*}

Another case of cosmological evolution corresponds to
$H\rightarrow 0$. This is another attractor of system of cosmological equations
which is realized when $\gamma <1$. Density of the holographic component
also approaches zero and $\rho _{de}/3H^{2}\rightarrow 0$. Universe expansion
asymptotically stops and for this scenario perturbations decrease with
time.

\begin{figure*}[t]
\includegraphics[scale=0.3]{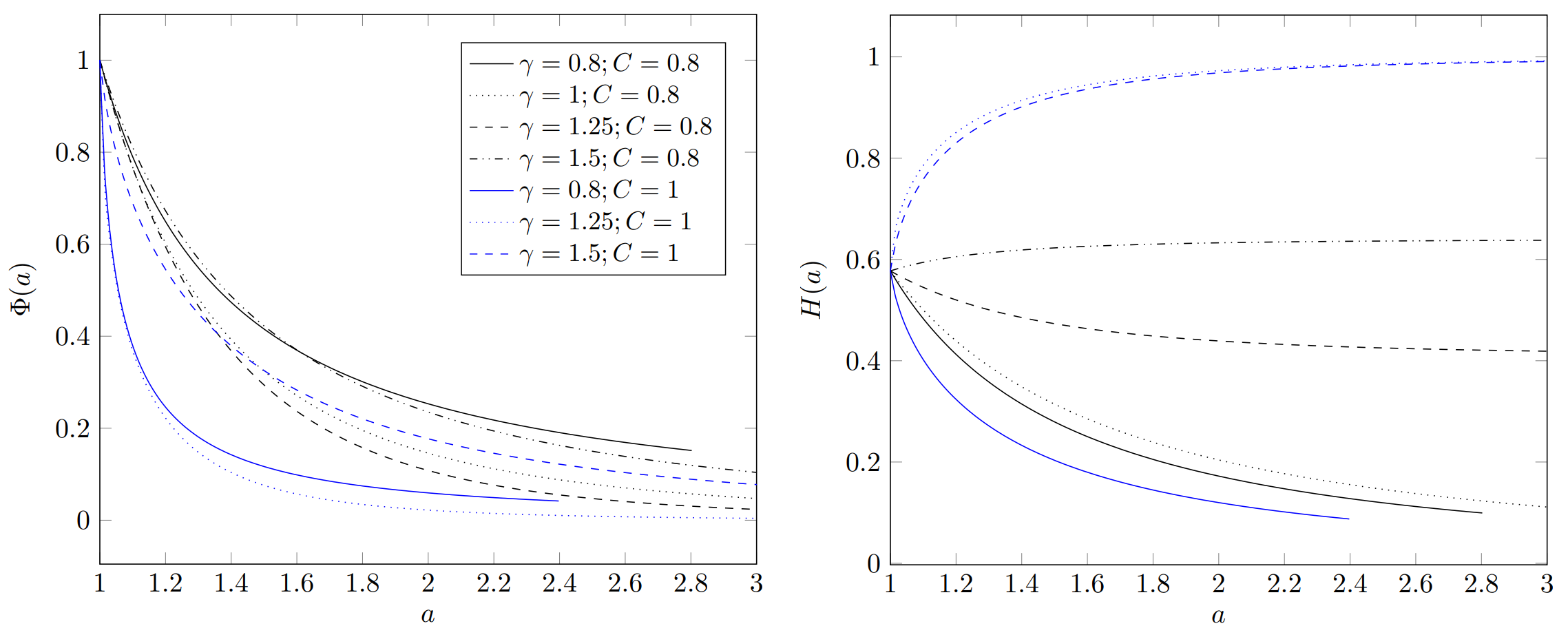}
\caption{Evolution of perturbations (left panel) and corresponding cosmological
dynamics (right panel) for some values of $\gamma $ and $C$. For all considered
values of $\gamma $ and $C$ perturbations decay with time.}
\label{fig10_v1}
\end{figure*}

We depicted the evolution of perturbations and corresponding dependence
of Hubble parameter from time on fig. \ref{fig10_v1}.

\section{Conclusion}
\label{sec7}

We investigated cosmological model of holographic dark energy in general
Tsallis model with $\rho _{de} \sim L^{2\gamma - 4}$ with two variants
of infrared cut-off $L$, event horizon and inverse Hubble parameter
$H^{-1}$. The main question of our paper is the evolution of metric perturbations
and matter perturbations for various values of model parameters. In classical
approach the criteria for stability is positiveness of square of sound
speed which calculates as $dp_{de}/d\rho _{de}$. But holographic dark energy
is caused by boundaries of universe and therefore we need to calculate
perturbations of horizon using corresponding definition for event horizon
or chosen scale cut-off.

In our analysis we considered firstly evolution of perturbations since
current moment of time when dark energy dominates, neglecting matter density
perturbations. Then we included in our consideration contribution of matter
density perturbations and consider evolution from past with very negligible
fraction of dark energy. Our investigation shows that account of matter
perturbations is very important for metric perturbations. In principle
one can say that evolution of metric perturbations defined by mainly by
matter perturbations. For wide interval of non-additivity parameter
$\gamma $ and $C$ perturbations of metric asymptotically approach constant.

We also investigated possible interaction between holographic component
and matter. For interaction
$\sim H(\alpha \rho _{m} + \beta \rho _{de})$ with realistic small values
of $\alpha $ and $\beta $ freezing of perturbations is faster than it would
be for non-interacting dark energy. Even in case of perturbation growth
from early times we have its smoothing in future. One note also that evolution
of matter perturbations in this case depends from parameters of integration
mainly parameter $\alpha $. For negative values of $\alpha $ the asymptotical
value of $\delta _{m}$ lies below in comparison with non-interacting case.

If inverse value of Hubble parameter is taken for infrared cut-off, equation
for perturbation becomes more simple. Universe filled with Tsallis HDE
with this cut-off can approach de Sitter regime on large times when
$H\rightarrow \text{const}$. For this we can expect that perturbations will
be smoothed out by cosmological expansion but decaying takes place only
for $\gamma >1$. For $\gamma <1$ there is a possibility that expansion
stops ($H\rightarrow 0$ for $t\rightarrow 0$). Perturbations asymptotically
decay for $H\rightarrow 0$. For quasi-de Sitter evolution with
$\gamma <1$ perturbations will increase.

Therefore we can conclude that analysis of perturbations for Tsallis HDE
based on consideration of HDE as boundary phenomena shows that these cosmological
models do not suffer from the problem of perturbations growth.

%


\begin{acknowledgments}
This research was supported by funds provided through the Russian Federal
Academic Leadership Program ``Priority 2030'' at the Immanuel Kant Baltic Federal University.
\end{acknowledgments}

%

\end{document}